\documentclass[twocolumn]{emulateapj}
\usepackage{epsfig}
\usepackage{amsmath}
\usepackage{graphicx}
\usepackage{natbib}
\usepackage{footnote}
\usepackage{amssymb}
\usepackage{multirow}
\usepackage{longtable}
\usepackage[hyperfootnotes=false]{hyperref}
\usepackage[hang, flushmargin]{footmisc}
\shorttitle{Structural Evolution of Early-type Galaxies to z=2.5 in CANDELS}
\shortauthors{Chang et al.}
\begin{document}
\title{Structural Evolution of Early-type Galaxies to z=2.5 in CANDELS}
\author{Yu-Yen~Chang\altaffilmark{1},
Arjen~van der Wel\altaffilmark{1},
Hans-Walter~Rix\altaffilmark{1},
Bradford~Holden\altaffilmark{2},
Eric F.~Bell\altaffilmark{3},
Elizabeth J.~McGrath\altaffilmark{4},
Stijn~Wuyts\altaffilmark{5},
Boris~H\"aussler\altaffilmark{6},
Marco~Barden\altaffilmark{7},
S. M.~Faber\altaffilmark{2},
Mark~Mozena\altaffilmark{2},
Henry C.~Ferguson\altaffilmark{8},
Yicheng~Guo\altaffilmark{2},
Audrey~Galametz\altaffilmark{9},
Norman A.~Grogin\altaffilmark{8},
Dale D.~Kocevski\altaffilmark{2},
Anton M.~Koekemoer\altaffilmark{8},
Avishai~Dekel\altaffilmark{10},
Kuang-Han~Huang \altaffilmark{8,11}, 
Nimish P.~Hathi \altaffilmark{12}, 
and Jennifer~Donley \altaffilmark{13} 
}
\newpage
\altaffiltext{1}{Max-Planck Institut f\"ur Astronomie, K\"onigstuhl 17, D-69117, Heidelberg, Germany; e-mail: chang@mpia.de}  
\altaffiltext{2}{UCO/Lick Observatory, Department of Astronomy and Astrophysics, University of California, Santa Cruz, CA 95064, USA}
\altaffiltext{3}{Department of Astronomy, University of Michigan, 500 Church Street, Ann Arbor, MI 48109, USA}
\altaffiltext{4}{Department of Physics and Astronomy, Colby College, Waterville, ME 04901, USA}
\altaffiltext{5}{Max-Planck-Institut f\"ur Extraterrestrische Physik, Postfach 1312, Giessenbachstr., D-85741 Garching, Germany}
\altaffiltext{6}{Schools of Physics \& Astronomy, University of Nottingham, University Park, Nottingham NG7 2RD, UK}
\altaffiltext{7}{Institute of Astro- and Particle Physics, University of Innsbruck, Technikerstra\ss e 25, A-6020 Innsbruck, Austria}
\altaffiltext{8}{Space Telescope Science Institute, 3700 San Martin Drive, Baltimore, MD 21218, USA}
\altaffiltext{9}{INAF - Osservatorio di Roma, Via Frascati 33, I-00040, Monteporzio, Italy}
\altaffiltext{10}{Racah Institute of Physics, The Hebrew University, Jerusalem 91904 Israel}
\altaffiltext{11}{Department of Physics and Astronomy, Johns Hopkins University, Baltimore, MD 21218, USA}
\altaffiltext{12}{Observatories of the Carnegie Institution for Science, Pasadena, CA, USA}
\altaffiltext{13}{Los Alamos National Laboratory, Los Alamos, NM, USA}
\newpage\null
\begin{abstract}
Projected axis ratio measurements of 880 early-type galaxies at redshifts $1<z<2.5$
selected from CANDELS are used to reconstruct and model their intrinsic shapes.
The sample is selected on the basis of multiple rest-frame colors to reflect
low star-formation activity. We demonstrate that these galaxies as an ensemble are dust-poor
and transparent and therefore likely have smooth light profiles, similar to visually classified early-type galaxies.
Similar to their present-day counterparts, the $z>1$ early-type galaxies show
a variety of intrinsic shapes; even at a fixed mass, the projected axis ratio 
distributions cannot be explained by the random projection of a set of galaxies with
very similar intrinsic shapes. However, a two-population model for the intrinsic shapes,
consisting of a triaxial, fairly round population, combined with a flat ($c/a\sim0.3$) oblate population,
adequately describes the projected axis ratio distributions of both present-day and
$z>1$ early-type galaxies. We find that the proportion of oblate versus triaxial galaxies
depends both on the galaxies' stellar mass, and - at a given mass - on redshift.
For present-day and $z<1$ early-type galaxies the oblate fraction strongly depends on galaxy mass.
At $z>1$ this trend is much weaker over the mass range explored here ($10^{10}<M_*/M_{\odot}<10^{11}$),
because the oblate fraction among massive ($M_*\sim10^{11} M_{\odot}$) was much higher in the past:
$0.59\pm 0.10$ at $z>1$, compared to $0.20\pm 0.02$ at $z\sim 0.1$. 
When combined with previous findings that the number density and sizes of early-type galaxies substantially increase over the same redshift range, this can be explained by the gradual emergence 
of merger-produced elliptical galaxies, at the expense of the destruction of pre-existing disks that were common among their high-redshift progenitors. In contrast, the oblate fraction among low-mass early-type galaxies ($\log(M_*/M_{\odot})<10.5$) increased toward the present, 
from $0.38\pm 0.11$ at $z>1$ to $0.72\pm 0.06$
at $z=0$. We speculate that this lower incidence of disks at early cosmic times can be attributed to two factors: low-mass, star-forming progenitors at $z>1$ were not settled into stable disks to the same degree as at later cosmic times, and the stripping of gas from star-forming disk galaxies in dense environments is an increasingly important process at lower redshifts.
\end{abstract}
\section{Introduction}
\label{sec1}
Early-type galaxies show a large variety in spatial and kinematic
structure \citep[e.g.,][and references
therein]{1996ApJ...464L.119K,2011MNRAS.414..888E}.  Among early types
with typical luminosities ($L^*$) or stellar masses, most have
disk-like properties in that they are axisymmetric, rotating and
intrinsically flat, even though their light profiles are significantly
more concentrated than those of late-type, star-forming $L_*$
galaxies.  More massive early-type galaxies are rounder, triaxial, and
slowly rotating.  Given these fundamental differences, one may surmise
that disk-like and spheroid-dominated galaxies have different
evolutionary paths and formation mechanisms.  Here we empirically
address this issue by analyzing the shape distribution of early-type
galaxies as a function of redshift.  Our reconstruction of the
internal structure of early-type galaxies at different cosmic epochs
will provide insight into the assembly history of massive, triaxial
galaxies as well as the evolutionary path of less massive, disk-like
early-type galaxies.

The internal structure of galaxies has been studied by means of analyzing projected shape distributions for several decades.  Early on, axisymmetric structure was assumed to describe the three-dimensional light profile of galaxies, that is, the projection of simple oblate and prolate models was used \citep{1926ApJ....64..321H, 1970ApJ...160..831S, 1978MNRAS.183..501B, 1983AJ.....88.1626F}. Then the triaxial model family \citep{1977ApJ...213..368S, 1985MNRAS.212..767B, 1991ApJ...383..112F} was considered to account for observational evidence that local early-type galaxies are not axisymmetric \citep{1992ApJ...396..445R, 1992MNRAS.258..404L, 1995AJ....110.1039T, 2007ApJ...670.1048K, 2008MNRAS.388.1321P, 2010A&A...521A..71M}. \citet{1996AJ....111.2243T} showed that the projected axis ratio distribution of early-type galaxies is accurately described by a model that consists of an oblate and a triaxial set of objects. Brighter galaxies tend to be more triaxial (non-axisymmetric) than fainter galaxies, which are more axisymmetric and intrinsically flatter \citep{2005ApJ...623..137V}.
This two-component model does not provide a mathematically unique
solution, but is physically plausible, in line with the kinematic
distinction of `fast rotators' and `slow rotators' \citep[e.g.,][]{2011MNRAS.414..888E}.

\citet[hereafter vdW09]{2009ApJ...706L.120V} used stellar masses instead of luminosity, and described the projected axis ratio distribution of early-type galaxies. In addition to enabling a more immediate comparison with galaxy formation models, the use of stellar masses instead of luminosities simplifies the interpretation of evolution with redshift \citep[][hereafter H12]{2012ApJ...749...96H} \citep[also see][]{2009ApJ...693..617H} vdW09 and H12 found that at all redshifts $z\lesssim1$ there is a quite sudden transition in the projected axis ratio distribution at a stellar mass of $\sim10^{11}M_\odot$. At lower masses the projected axis ratio distribution is broad, indicative of a large fraction of disk-like early-type galaxies, which have a ceiling mass of $\sim2\times10^{11}M_\odot$, above which essentially all early-type galaxies are intrinsically round. H12 provide a quantitative analysis by describing the projected axis ratio distribution of early-type galaxies, and its evolution with redshift by the aforementioned two-component model. Overall, they found little evolution between $z=0.8$ and the present. \citet{2011ApJ...730...38V} and \citet[hereafter C13]{2013ApJ...762...83C} extended these studies to higher redshift. They found that massive early-type galaxies at $z\gtrsim 1.5$ are flatter than at the present. Their implied disk-like structures show that these galaxies formed while gas had time to settle into disks. 

The vdW09, H12, and C13 samples were selected by (a lack of) star formation activity \citep[also see][]{2007ApJ...655...51W,2009ApJ...691.1879W}.  Such a selection can effectively be used as a proxy for a (visual) morphological classification, as a smooth light profile is the main criterion for the visual classification of an early-type galaxy, which usually corresponds to low star-formation activity \citep[also see][]{2012ApJ...748L..27P}.  A practical advantage of a star formation selection is that it allows for the consistent selection of high-redshift samples, for which visual classification is difficult or impossible.  Furthermore, since we are investigating the evolution of structural properties, the use of structural parameters such as concentration or S\'ersic index to select early-type galaxies are prone to introducing biases.

So far, these results have been rather qualitative. In this paper, we provide a more quantitative description of the internal structure of $z=1-2.5$ early-type galaxies down to $M\sim10^{10}$.
The Cosmic Assembly Near-infrared Deep Extragalactic Legacy Survey \citep[CANDELS, ][]{2011ApJS..197...35G,2011ApJS..197...36K}, a 902 orbit Hubble Space Telescope (HST) multi-cycle treasury program, provides high-resolution near-infrared imaging aimed at investigating the structural and morphological properties of galaxies to $z\sim3$ in the rest-frame optical. \citet[hereafter vdW12]{2012ApJS..203...24V} used \texttt{GALFIT} \citep{2010AJ....139.2097P} to measure the global structural parameters of $\sim$ 100,000 galaxies in CANDELS. We draw from this work to construct a sample of 569 $z>1$ early-type galaxies with accurately measured axis ratios.

The structure of this paper is as follows. In Section~\ref{sec2} we describe the data and select our sample of early-type galaxies. 
In Section~\ref{sec3} we analyze the structural parameters of early-type galaxies and their evolution since $z\sim2.5$ and as a function of stellar mass. In Section~\ref{sec4} we describe our models to reconstruct the intrinsic shape distribution. In Section~\ref{sec5} we investigate the internal structure of early-type galaxies and its evolution. In Section~\ref{sec6} and \ref{sec7} we discuss and summarize our results.

We use AB magnitudes and adopt the cosmological parameters ($\Omega_M$,$\Omega_\Lambda$,$h$)=(0.27,0.73,0.70) in this paper.

\section{Data}
\label{sec2}

\begin{figure}[ht]
\centering
\includegraphics[width=1.0\columnwidth]{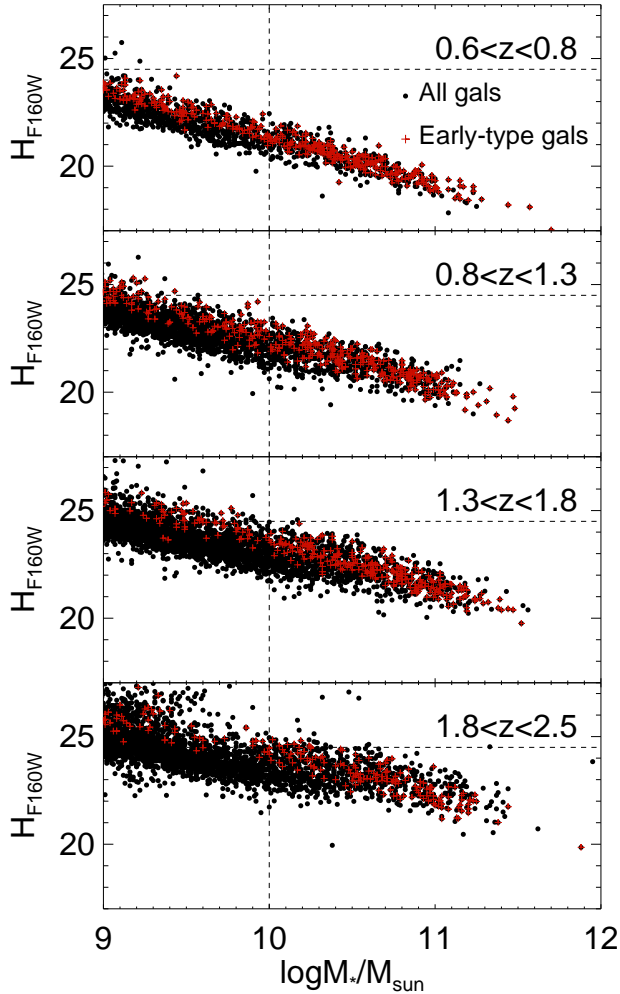}
\caption{CANDELS $H_{F160W}$ magnitude vs. stellar mass at different redshifts. The red crosses represent early-type galaxies, selected as described in Section~\ref{sec2_3} and illustrated in Figure~\ref{ca_fig_uvj}.  The black symbols represent all galaxies. We adopt $H_{F160W}=24.5$ as our magnitude limit: vdW12 showed that size and shape measurements are better than 10\% down to this limit.  This leads us to adopt a stellar mass limit of $\log(M_*/M_{\odot})>10$, ensuring robust structural parameter estimates for all galaxies in our sample up to $z\sim 2.5$.}
\label{ca_fig_m_h}
\end{figure}

\begin{figure}[ht]
\centering
\includegraphics[width=1.0\columnwidth]{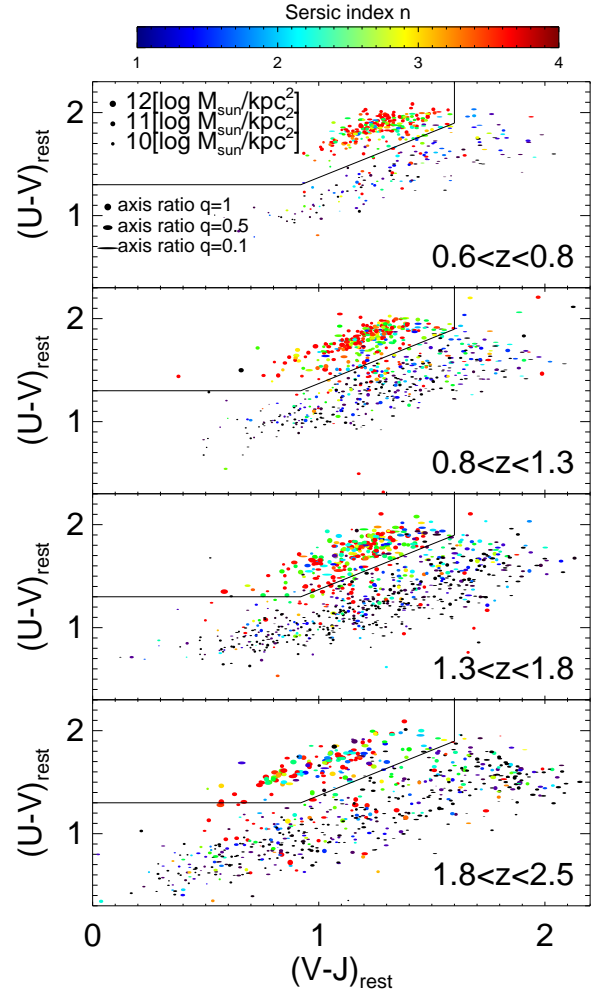}
\caption{Rest-frame $U-V$ vs.~rest-frame $V-J$ colors for galaxies in four redshift bins.  As shown, e.g., by \citet{2009ApJ...691.1879W}, the black polygons effectively separate star-forming and quiescent galaxies, which we use in this paper to select our early-type sample.  The symbols' color coding corresponds to the S\'ersic index $n$, the symbol size (area) with stellar mass surface density ($M_*/2\pi q R_{\rm{eff}}^2$), and symbol shape with observed, projected axis ratio $q$. The color-color selection separation of early- and
late-type galaxies corresponds well with their structural properties
in the sense that early-type galaxies have high S\'ersic indices and
large surface mass densities.} 
\label{ca_fig_uvj}
\end{figure}

\subsection{Multi-wavelength Data and SED Fitting}
\label{sec2_1}
In this paper, we use imaging and multi-wavelength catalogs from CANDELS in  
the Great Observatories Origins Deep Survey-South field 
\citep[GOODS-S, ][`wide' over 4'$\times$10' and `deep' over 7$\times$10']{2004ApJ...600L..93G}
and Ultra Deep Survey field \citep[UDS; ][‘wide’ over 9'$\times$24']{2007MNRAS.379.1599L}.
The deep near-infrared HST survey allows us to select early-type galaxies up to $z=2.5$.
The details of the multi-wavelength catalogs are described in Guo et al., (2013, GOODS-S), \citet[][UDS]{2013ApJS..206...10G}, and \citet[][IRAC SEDS catalog]{2013ApJ...769...80A}.

The method and algorithms for acquiring photometric redshifts, rest-frame colors and stellar masses
are described by \citet{2011ApJ...742...96W}.
Briefly, photometric redshifts are estimated 
by EAzY \citep{2008ApJ...686.1503B} and available
spectroscopic redshifts are included.
The stellar masses, star-formation rates, and rest-frame colors
are estimated by FAST \citep{2009ApJ...700..221K}. 
The \citet{2003MNRAS.344.1000B} model, 
and a \citet{2003PASP..115..763C} stellar initial mass function is adopted.
A range of ages, star formation histories and extinction
parameters is explored.

\subsection{Galaxy Structural Parameters}
\label{sec2_2}
The structural parameters (radii, S\'ersic indices and projected axis ratios) are taken from vdW12 who fit single S\'ersic profiles to individual galaxies with \texttt{GALFIT}. 
Many of the galaxies in our sample are very small ($\sim$ 1kpc), close to the resolution limit. If the point spread function (PSF) is precisely known, this is not a problem as shown by vdW12, at least under the assumption that the characterization of the light profile by a single S\'ersic component is reasonable. In order to test the sensitivity of our results to errors in the PSF model, we refit our sample with the `wrong' PSF: if we convolve the S\'ersic profile with the F125W PSF model in order to fit the F160W images, the resulting axis ratios are larger, but not to the extent that our results are affected. Since we know the F160W PSF with much better accuracy than the $\sim 15\%$ difference between the F125W and F160W PSFs (FWHM$_{F125W}\sim0''.20$; FWHM$_{F160W}\sim0''.17$), we conclude that errors in our PSF model do not affect our results.

\subsection{Sample Selection}
 \label{sec2_3}
Combining the multi-wavelength and structure parameter catalogs, we have an initial sample of 56,010 objects (21,889 in GOODS-S and 34,121 in UDS).
Size and shape measurements are accurate and precise to 10\% for galaxies with $H_{F160W}\sim24.5$ (see vdW12).
We adopt a stellar mass limit of $M=10^{10}M_\odot$, which allows us to consistently compare galaxies at all redshift $z<2.5$ (see  Figure~\ref{ca_fig_m_h}). We reject stars by including only objects with $J-H>0.15$.  We only include galaxies with good {\texttt GALFIT} fits (flag=0; 87\% of the remaining sample) from the vdW12 catalog and ignore 13\% with suspect fits (flag=1) or bad fits (flag=2).  This mass-selected sample with reliable structure measurements consists of 2,827 objects.

\setcounter{footnote}{0}

To separate quiescent galaxies from star-forming galaxies, we use color-color selection criteria as shown in Figure~\ref{ca_fig_uvj}, 
following: $(U-V)>0.88\times(V-J)+0.49$, $(U-V)>1.3$ and $(V-J)<1.6$. 
\footnote{To compute the rest-frame $U$, $V$, and $J$ band fluxes, we use the $UX$ and $V$ Bessell filters and the Palomar $J$ filter.}
We define these as early-type galaxies and the remainder as late-types. 
This approach follows the technique outlined by, \citet[e.g.,][]{2009ApJ...691.1879W}, 
but the color selection criteria are slightly different to account for differences 
in filter transmission curves and small offsets in the flux measurements.
In Figure~\ref{ca_fig_uvj} it can be seen that this star-formation activity- based selection corresponds well with the S\'ersic index, indicating that our selection by star formation activity is effectively equivalent to a concentration-based definition of early type \citep[also see][]{2008ApJ...682..355B,2011ApJ...742...96W,2012ApJ...753..167B,2012ApJ...753..114W} over the full redshift range probed here. As noted before, star-formation activity is strongly anti-correlated with S\'ersic index and surface mass density, up to at least $z=2.5$.
Even though in this paper we emphasize the diskiness 
of early-type galaxies,  it is also apparent in Figure~\ref{ca_fig_uvj} that 
late-type galaxies are still flatter, that is, more disk-like, than early-type galaxies at all redshifts.

The final sample of mass-selected early-type galaxies with reliable (flag=0) structure measurements consists of 880 galaxies in the redshift range $0.6<z<2.5$.
The numbers of galaxies in different redshift bins are shown in Table~\ref{number}.
We create three stellar mass bins for CANDELS with a roughly equal number of galaxies.

\begin{table}[ht]
\caption{Sample Sizes} 
\label{number}
\centering
\begin{tabular}{|c|c|ccc|}
\hline 
$\log(M_*/M_\odot)$  & $10.1-11.5$ & $10.8-11.5$& $10.5-10.8$ & $10.1-10.5$ \\
\hline
Redshift  &  \multicolumn{4}{c|}{Numbers} \\
\hline 
SDSS & 32842 & 13640 & 13991 & 5211 \\
H12 & 1321 & 384 & 475 & 462\\
$1<z<2.5$ & 569 & 197 & 168 & 204 \\
$0.6<z<0.8$ & 220 & 47 & 67 & 106 \\
$0.8<z<1.3$ & 256 & 78 & 66 & 112 \\
$1.3<z<1.8$ & 244 & 88 & 71 & 85 \\
$1.8<z<2.5$ & 147 & 55 & 47 & 45 \\
\hline
\end{tabular}
\end{table}

The SDSS sample from H12 is used as a low-redshift benchmark. Here, the early-type galaxies are selected by an equivalent color-color criterion.
We verified that rejecting all SDSS color-color selected early-type
galaxies with detected $H\alpha$ emission ($\sim 18\%$ of the sample)
does not change our results. Even though the galaxies with detected
emission lines are on average somewhat flatter than their counterparts
without emission lines, the axis ratio distribution analyzed in the
subsequent sections is not significantly altered.

Sufficiently deep emission-line data are not available for the
galaxies in CANDELS. Instead, we search for detections in public MIPS
24$\mu$m imaging in the
UDS\footnote{http://irsa.ipac.caltech.edu/data/SPITZER/SpUDS} and
cataloged MIPS 24$\mu$m flux measurements from \citet{2008ApJ...682..985W} in
GOODS-S.  Removing the $3\sigma$ detected objects ($\sim 16\%$) does
not change the projected axis ratio distributions significantly.  We
conclude that the evolutionary trends with redshift are not sensitive
to the inclusion of contaminating populations of star-forming galaxies
and/or active galactic nuclei.

\begin{figure}[ht]
\centering
\includegraphics[width=1.0\columnwidth]{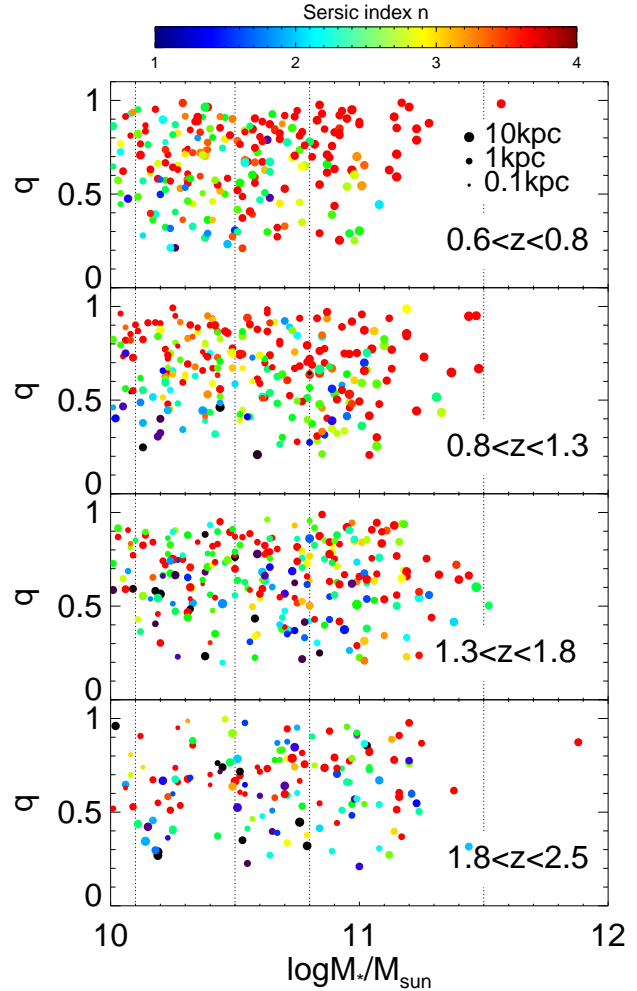}
\caption{Projected axis ratio vs. stellar mass for early-type galaxies in CANDELS in four redshifts bins. The symbols' color coding corresponds to the S\'ersic index $n$, the symbol size represents the radius in kiloparsecs. High-mass early-type galaxies are rounder and have higher S\'ersic indices than low-mass early-type galaxies, but these trends apparently weaken at $z\sim 2$.  At all redshifts, flatter galaxies have lower S\'ersic indices, indicating that the population exists of a mix of different types of galaxies, and that variation in projected shape is not only the result of different viewing angles.  The thin vertical lines indicate the mass bins that we use in this paper,} and are chosen to contain similar numbers of galaxies.
\label{ca_fig_m_q}
\end{figure}
\begin{figure}[ht]
\centering
\includegraphics[width=1.0\columnwidth]{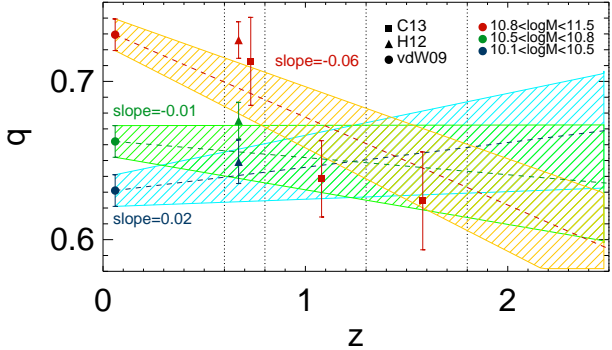}
\caption{Evolution of the projected axis ratios for galaxies in three mass bins (color coding). The filled symbols are median values of previously published results.  The dashed lines represent fits to the redshift-axis ratio distribution of the individual objects in the CANDELS sample, anchoring to the median axis ratio of present-day galaxies.  The shaded area indicates the $1-\sigma$ uncertainty. See the text for details on the fitting method.  The most significant evidence for evolution is seen for the massive galaxies: these are progressively flatter at higher redshift.}
\label{ca_fig_z_evolution}
\end{figure}
\begin{figure*}[ht]
\centering
\includegraphics[width=0.9\textwidth]{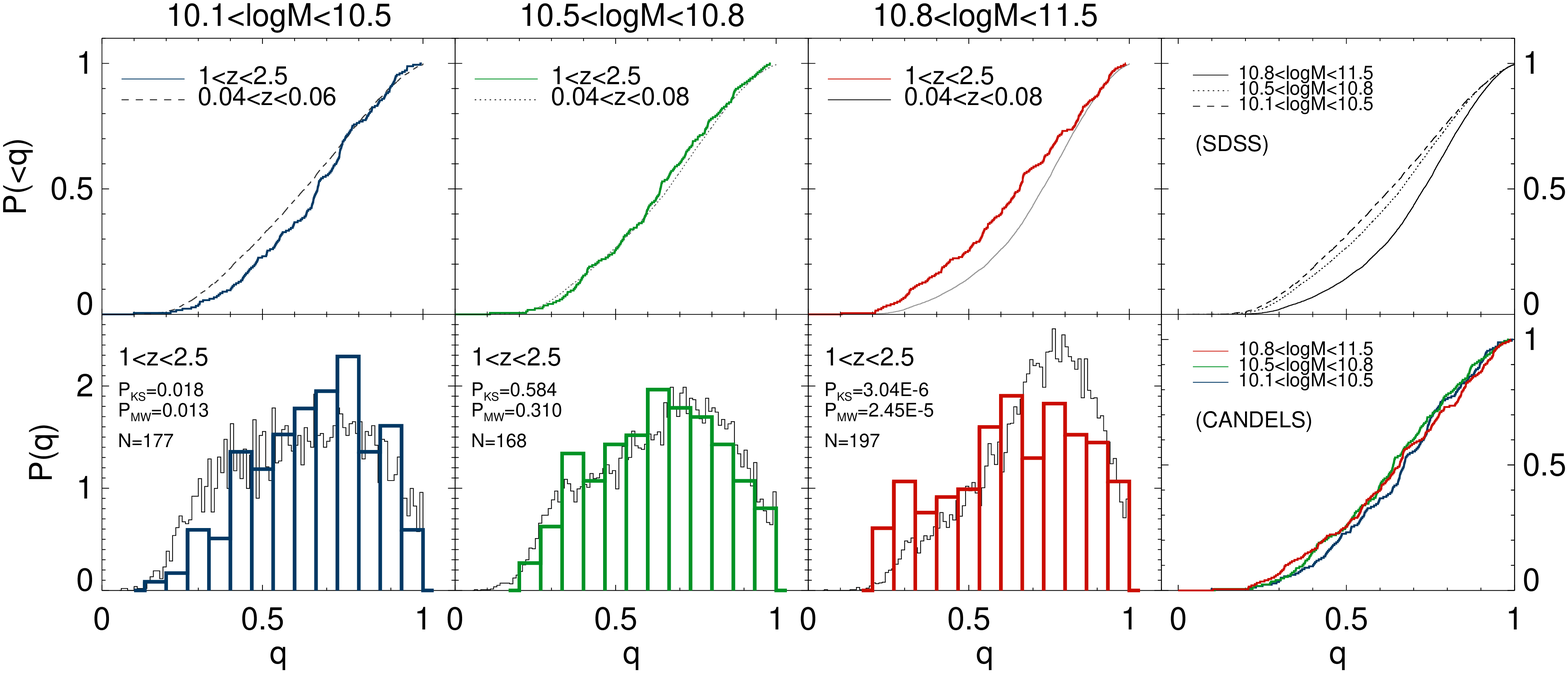}
\caption{Redshift evolution of the projected axis ratio distributions in different mass bins.  The top row of panels shows cumulative distributions; the bottom row shows binned histograms.  The colored lines and histograms represent the $1<z<2.5$ CANDELS early-type galaxy sample; the gray/black lines represent the present-day early-type galaxy sample from SDSS.  The right most panels combine the cumulative histograms for the present-day sample (top) and high-redshift sample (bottom), separated according to the mass bins. Probabilities that the distributions are consistent with each other from two statistical tests (Kolmogorov-Smirnov and Mann-Whithney) are given in the bottom panels.  The samples high- and low-mass bins show evidence for evolution in opposite directions: whereas high-mass galaxies are flatter at high redshifts, low-mass galaxies are rounder.}
\label{ca_fig_q_hist}
\end{figure*}
\section{Evolution of the projected axis ratio distribution}
\label{sec3}
In Figure~\ref{ca_fig_m_q} we show the axis-ratio distributions of early-type galaxies as a function of stellar mass for a number of redshift bins from $z=0.6$ to $z=2.5$. Half-light radii ($R_{eff}$) and S\'ersic indices ($n$) are represented by varying the symbol size and color coding, respectively.  Over the whole stellar mass range probed here the typical $R_{eff}$ and $n$ increase from $z\sim2.5$ to later times, while non-starforming galaxies with exponential light profiles are rare at all redshifts.  As previously reported by vdW09, H12, and C13, the most massive galaxies are the roundest, which can be seen here in particular at $z\sim 1$; at higher redshifts the probed volume is too small to include a sufficiently large number of very massive galaxies.

The main results presented in this paper can all be qualitatively seen in Figure~\ref{ca_fig_m_q}.  First, as was also shown by C13, there are many flat early-type galaxies with mass $\sim 10^{11} M_{\odot}$ at $z>1$. Second, and contrarily, there are not many flat early-type galaxies with mass $\sim 10^{10} M_{\odot}$ at $z>1$.  The overall tendency is that the dependence of shape on galaxy mass is weak at $z>1$ and strong at $z<1$. To investigate these indications of structural evolution in a quantitative way, we will model the projected axis ratio distributions to infer the intrinsic shape distribution in Section~\ref{sec5}. 

But first, we will establish the significance of these trends in a model-independent manner. In Figure~\ref{ca_fig_z_evolution}, we perform least-squares fits to the axis ratios of the full $0.6<z<2.5$ sample separated into three mass bins, anchored by the low-redshift median values from the SDSS sample to which we assign a 0.01 systematic uncertainty (see H12).  The uncertainties on the least-square fits are obtained by bootstrapping the sample and perturbing the photometric redshift ($z_{phot}$) and the projected axis ratio by their measurement uncertainties. Moreover, uncertainties in stellar masses ($M_*$) are included in two steps: first, the perturbation in photometric redshift is propagated ($M_* \propto  (1+z)^4$) and, second, a random mass uncertainty of 0.2 dex \cite[see, e.g., ][]{2006ApJ...652...97V} is included. Figure~\ref{ca_fig_z_evolution} shows that there is significant evolution in the projected axis ratios for massive galaxies, with the projected axis ratios decreasing toward high redshift, and we find marginal evidence for increasing projected axis ratios with redshift for the lowest-mass sample. 

We now turn to the full distribution of axis ratios, which, compared to evolution in the average or median, enables more sensitive tests for structural evolution. In Figure~\ref{ca_fig_q_hist}, we compare the axis ratio distributions
of our $1<z<2.5$ early-type galaxies with local early-type galaxies
(see vdW09 and H12) by means of cumulative distributions and of histograms. Figure~\ref{ca_fig_q_hist} shows that for $\log(M_*/M_{\odot})>10.8$, high-redshift
galaxies are flatter (have smaller projected axis ratios) than local
galaxies, while for $\log(M_*/M_{\odot})<10.5$, high-redshift galaxies
are rounder. We use Kolmogorov-Smirnov (K-S) and Mann-Whitney U (M-W)
tests to show that these trends are significant at the $5\sigma$ and
$3\sigma$ level, respectively.  These quantitative comparisons confirm
the hints seen in Figure~\ref{ca_fig_m_q}.

While the flattening of high-mass galaxies is consistent with previous
results \citep[][, and C13]{2011ApJ...730...38V,2012MNRAS.427.1666B,2012ApJ...745..179W,2013MNRAS.428.1460B},
the 3$\sigma$-level evidence that low-mass early types were rounder at earlier epochs is surprising. One could 
suspect that systematic shape measurement errors may prevent us
from recovering the actual flatness of the small, faint galaxies in
this sub-sample.  However, the simulations performed by vdW12 indicate
that shapes and sizes can be recovered with high accuracy down to the
regime probed here.  Note, however, that those
simulations were performed with ideal S\'ersic profiles, not with real
galaxy light profiles.  In addition, we can ask whether
mismatches in the PSF model matter.  In
order to test this we rerun the profile fits on the F125W images of
this sub-sample of low-mass early-type galaxies. For this test 
we replace the F125W PSF model, which we assume to be accurate, 
with the F160W PSF model.  We know that the F160W PSF model is too
broad to describe the light profiles of point sources in the F125W
imaging (by $\sim 15\%$).  Therefore, the projected axis ratio will
now be underestimated (objects will appear flatter than they are).
Even with this crudely wrong PSF model we find that the axis ratios of
the low-mass $z>1$ early types are not flatter than the axis
ratios of their present-day counterparts.  Given that the uncertainty
in our PSF models is much smaller than the difference between the
F125W and F160W PSF models, we can safely conclude that the observed
evolution in the axis ratio distribution for low-mass early-type
galaxies is not due to uncertainties in our PSF models.

Now we proceed in the next two sections to reconstruct the intrinsic structural properties of early-type galaxies as a function of stellar mass and redshift. We explore a variety of approaches that employ different model families and search for solutions by assuming random viewing angle distributions for our samples. In Section~\ref{sec4} we apply an analytical approximation to reconstruct the intrinsic axis ratio distribution of axisymmetric model families. In Section~\ref{sec5} we project model distributions that represent a combination of axisymmetric and triaxial families in order to reproduce the observed distributions of projected axis ratios and to find best-fitting solutions.
\begin{figure*}[ht]
\centering
\includegraphics[width=1.0\textwidth]{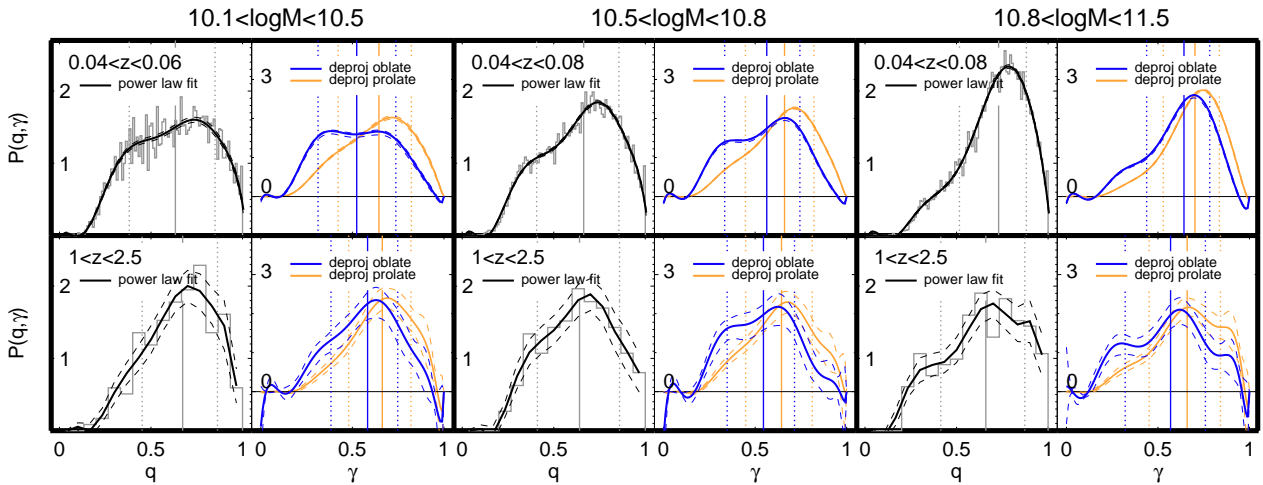}
\caption{Projected axis ratio ($q$) distributions in gray/black and
deprojected, intrinsic axis ratio ($\gamma$) distributions in
blue/orange, inferred as described in Section~\ref{sec4}. Observed, projected axis ratio distributions (black histograms) are represented by eighth-order polynomials (black lines) and then analytically deprojected according to Equation~(\ref{deproj_1}) to infer the intrinsic shape distribution of axisymmetric model populations (oblate in blue; prolate in orange). The dashed lines below and above the solid lines represent the 16 and 84 percentile confidence intervals obtained from bootstrapping
($n=10,000$).  The vertical lines show the 16, 50, and 84 percentiles
of the deprojected intrinsic axis ratios.  The top row of panels shows present-day early-type galaxies from SDSS; the bottom row shows $1<z<2.5$ early-type galaxies from CANDELS.}
\label{ca_fig_deprojections}
\end{figure*}
\section{Analytical Reconstruction of the Intrinsic Shape Distribution}
\label{sec4}
For an oblate ellipsoid at the origin of a Cartesian coordinate system, the intrinsic shape can be written as $x^2+y^2+z^2/\gamma^2=1$, 
where $\gamma$ ($0<\gamma\leq 1$) is the intrinsic axis ratio between the (one) short axis and the (two) long axes. For a prolate ellipsoid, the intrinsic shape can be written as 
$x^2/\gamma^2+y^2/\gamma^2+z^2=1$, where $\gamma$ ($0<\gamma\leq 1$) is the intrinsic axis ratio between the (two) short axes and the (one) long axis.
The intrinsic axis-ratio distribution, $\psi(\gamma)$, can be inferred as prescribed by \citet{1983AJ.....88.1626F}:
\begin{subequations}
\begin{align}
& \psi_O(\gamma)=\frac{2}{\pi}\sqrt{1-\gamma^2}\frac{d}{d\gamma}\int^q_0
\frac{\phi_O(q)dq}{\sqrt{\gamma^2-q^2}} \\
& \psi_P(\gamma)=\frac{2}{\pi}\frac{\sqrt{1-\gamma^2}}{\gamma^2}\frac{d}{d\gamma}\int^q_0  
\frac{\phi_P(q)q^3 d q}{\sqrt{\gamma^2-q^2}},
\end{align}
\label{deproj_1}
\end{subequations}
where $\phi$ is the projected axis-ratio distribution, and 
the subscripts $O$ and $P$ refer to the oblate and prolate case, respectively.
If we describe the projected axis-ratio distribution by a power law ($\phi(q)=(m+1)q^m$ with $m>-1$), we can rewrite 
Equation~(\ref{deproj_1}) analytically:
\begin{subequations}
\begin{align}
& \psi_O(\gamma)=\frac{2\gamma^{m-1}\sqrt{1-\gamma^2}}{B(0.5m,1.5)} \\
& \psi_P(\gamma)=\frac{2\gamma^{m}\sqrt{1-\gamma^2}}{B(0.5m+0.5,1.5)},
\end{align}
where $B(x,y)$ is the beta function.
\label{deproj_2}
\end{subequations}
The reconstructed intrinsic axis ratio distribution should be non-negative if an oblate or prolate model is a good description of the data.
\subsection{Application}
\label{sec4_1}
Figure~\ref{ca_fig_deprojections} shows the results of the deprojection outlined above.  We use an eighth-order power law, $\phi(q)=\Sigma^{8}_{m=0} C_m (m+1)q^m$, to describe the observed projected axis ratio distributions (black lines in Figure~\ref{ca_fig_deprojections}). 
The dashed lines show the 16 and 84 percentile confidence intervals
obtained from bootstrapping \citep[e.g.,][]{1995AJ....110.1039T,1996AJ....111.2243T,1996ApJ...461..146R,1996ApJ...471..822R}.  
The reconstructed intrinsic shape distributions for
the oblate and prolate models (shown in thick blue and orange lines,
respectively) are sometimes slightly negative, but the uncertainties
are such that this can be attributed to the limited sample size.
The distributions are very broad; that is, in narrow ranges of mass, galaxies display a large variety in intrinsic shape, and the population cannot consist of objects that are all similar in intrinsic thickness.  This is true both for present-day galaxies and for $z>1$ galaxies. Changes in the intrinsic shape distribution with redshift mirror changes in the projected shape distribution: high-mass galaxies were on average flatter at $z>1$, and low-mass galaxies were rounder.  Especially for the large, present-day samples, there is a clear hint that multiple components (galaxy populations) are needed to describe the intrinsic shape distribution, which we will explore below.
\section{Projection of Axisymmetric and Triaxial Models}
\label{sec5}
Following \citet{1985MNRAS.212..767B} Equation (11) and (12), we project a triaxial ellipsoid (written as $m^2=x^2/a^2+y^2/b^2+z^2/c^2$, at the origin of Cartesian coordinate system) and compute the projected axis ratio $q$ as follows:
\begin{subequations}
\begin{align}
& A=\frac{\cos^2\theta}{\gamma^2}\Biggl(\sin^2\phi+\frac{\cos^2\phi}{\beta^2}\Biggl)+\frac{\sin^2\theta}{\beta^2} \\
& B=\cos\theta\sin2\phi\Biggl(1-\frac{1}{\beta^2}\Biggl)\frac{1}{\gamma^2} \\
& C=\Biggl(\frac{\sin^2\phi}{\beta^2}+\cos^2\phi\Biggl)\frac{1}{\gamma^2} \\
& q(\theta,\phi;\beta,\gamma)=\sqrt{\frac{A+C-\sqrt{(A-C)^2+B^2}}{A+C+\sqrt{(A-C)^2+B^2}}},
\end{align}
\label{proj_1}
where ($\theta$,$\phi)$ are the polar and azimuthal viewing angles in
a spherical coordinate system, and $\beta=b/a$ and $\gamma=c/a$.  Note
that $\beta=1$ and $\beta=\gamma$ correspond to the special,
axisymmetric cases (oblate and prolate, respectively).
\end{subequations}
In order to account for variations in intrinsic shape, we assume a Gaussian
distribution for the triaxiallity $T$($=[1-\beta^2]/[1-\gamma^2]$) and
ellipticity $E$($=1-\gamma$) with dispersion $\sigma_T$ and
$\sigma_E$.

For a given set of parameters ($T$,$E$,$\sigma_T$,$\sigma_E$), we
numerically generate distributions for $\beta$ and $\gamma$.  Then, a
random viewing angle ($\theta$,$\phi$) is assigned to each of the
elements of the distribution (100,000 in our case) such that with
Equation~(\ref{proj_1}) the projected axis ratio distribution can be
generated.  This distribution corresponds to the probability
distribution $p(q_{model})$.

For nearly round ($q\sim1$) galaxies, random noise will always
cause the measured $q$ to be an underestimate as the position
angle of the long axis becomes ill-determined. This affects the
projected axis-ratio distribution as described by
\citet{1995ApJ...447...82R} (Equation (C5)):
 \begin{equation}
P_e(\epsilon,\epsilon_e,\Delta\epsilon)=\frac{\epsilon}{\Delta\epsilon^2}
I_o\biggl(\frac{\epsilon\epsilon_e}{\Delta\epsilon^2}\biggl)
\exp\biggl(-\frac{\epsilon^2+\epsilon_e^2}{2\Delta\epsilon^2}\biggl),
\label{proj_2}
\end{equation}
where $\epsilon$($=1-q$) is the measured ellipticity,
$\epsilon_e$ is the expected ellipticity, $\Delta\epsilon$ is the
measured error, and $P_e$ is the expected ellipticity distribution.
We numerically implement the difference between $\epsilon$ and
$\epsilon_e$ to correct the generated probability distribution
$p(q_{model})$.  We adopt fixed values $\Delta\epsilon=\Delta
q$ for each of the data sets used here: 0.03 and 0.05 for the
low- and high-$z$ data sets from H12, and 0.04 for the CANDELS data set.

The total likelihood $L$ for a measured set projected axis ratios
$q_{data}$ and a given set of model parameters $T$, $E$, $\sigma_T$,
and $\sigma_E$ is given by $L=\Sigma_{q_{data}} \log
p(q_{data}|q_{model}) $, where $p$ has a minimum value of 0.01.

$L$ is computed for a grid of model parameters, chosen in various ways
for the different approaches explored below, such that the maximum
likelihood model can be located in the grid and the best-fitting model
is identified.

In order to obtain uncertainty estimates on the best-fitting model
parameters, we bootstrap the observed data ($q_{data}$), also
perturbing $q_{data}$ by the measurement uncertainty and perturbing
the redshift and stellar mass estimates as described in Section~\ref{sec3}.
\begin{figure}[ht]
\centering
\includegraphics[width=1.0\columnwidth]{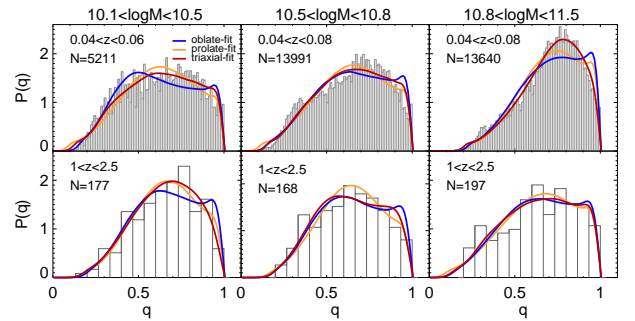}
\caption{Histograms show observed distributions of projected axis ratios for present-day early-type galaxies from SDSS (upper row) and at $1<z<2.5$ from CANDELS (bottom row), each in three mass bins.  The colored lines represent the best-fitting, single-component models with Gaussian distributions for intrinsic axis ratios, with the oblate model in blue, the prolate model in orange, and the triaxial model in red. See Section~\ref{sec5_1_1} for details.  The mean and dispersion of the best-fitting Gaussians are listed in Table \ref{model_single}.}  
\label{ca_fig_triaxial}
\end{figure}

\subsection{Application}
\label{sec5_1}
\subsubsection{Single-component Model for the Intrinsic Shape}
\label{sec5_1_1}

\begin{table*}[ht]
\caption{Single-component Fitting Results}
\label{model_single}
\centering
\begin{tabular}{|ccc|cccc|cc|}
\hline 
Model &  Mass($\log(M_*/M_\odot)$) & Redshift(z) & $T$ \footnote{$T$ is the mean triaxiality parameter, with standard deviation $\sigma_{\rm{T}}$; these are set to 0 or 1 for the oblate and prolate models.} & $\sigma_T$ & $E$ \footnote{$E$ and $\sigma_{E}$ are the ellipticity (1 minus the intrinsic short-long axis ratio) and its standard deviation.} & $\sigma_E$ & $P_{KS}$ \footnote{The final two columns list the K-S and M-W probabilities that the observed and best-fitting model projected axis ratio distributions are indistinguishable, for a randomly drawn realization of the model distribution with the same number of objects as the observed samples.  These serve as a crude goodness-of-fit test.} & $P_{MW}$ \\
\hline 
\multicolumn{9}{|c|}{Single oblate model} \\
\hline
Oblate & 10.8-11.5 & 0.04-0.08 (SDSS) & 0 & 0 & $0.48\pm0.01$ \footnote{Uncertainties are obtained from bootstrapping.} & $0.18\pm0.01$ & *0.00* \footnote{The asterisks (*) represent the significant probability is smaller than 5\%. It implies that the distributions are distinguishable.}& *0.01*\\
Oblate & 10.5-10.8 & 0.04-0.08 (SDSS) & 0 & 0 &  $0.61\pm0.00$ & $0.18\pm0.00$ & *0.00* & *0.01*\\
Oblate & 10.1-10.5 &  0.04-0.06 (SDSS) & 0 & 0 & $0.66\pm0.01$ & $0.13\pm0.01$ & *0.00* & 0.16\\
\hline
Oblate & 10.8-11.5 & 0.6-0.8 (H12) & 0 & 0 & $0.46\pm0.03$ & $0.17\pm0.03$  & 0.21 & 0.20\\
Oblate & 10.5-10.8 & 0.6-0.8 (H12) & 0 & 0 & $0.58\pm0.04$ & $0.17\pm0.04$ & 0.24 & 0.46\\
Oblate & 10.1-10.5 & 0.6-0.8 (H12) & 0 & 0 & $0.58\pm0.03$ & $0.14\pm0.03$  & 0.12 & 0.13\\
\hline
Oblate & 10.8-11.5 & 1-2.5 & 0 & 0 & $0.61\pm0.03$ & $0.18\pm0.03$ & 0.99 & 0.45\\
Oblate & 10.5-10.8 & 1-2.5 & 0 & 0 & $0.59\pm0.04$ & $0.14\pm0.03$  & 0.69 & 0.26\\
Oblate & 10.1-10.5 &  1-2.5 &  0 & 0 & $0.56\pm0.03$ & $0.15\pm0.03$ &  0.41 & 0.34\\
\hline 
\multicolumn{9}{|c|}{Single prolate model} \\
\hline
Prolate & 10.8-11.5 & 0.04-0.08 (SDSS) & 1 & 0 & $0.37\pm0.00$ & $0.18\pm0.00$ & *0.00* & 0.34\\
Prolate & 10.5-10.8 & 0.04-0.08 (SDSS) & 1 & 0 & $0.45\pm0.00$ & $0.20\pm0.01$ & *0.00* & 0.07\\
Prolate & 10.1-10.5 &  0.04-0.06 (SDSS) & 1 & 0 & $0.47\pm0.00$ & $0.20\pm0.00$ & *0.00* & 0.35\\
\hline
Prolate & 10.8-11.5 & 0.6-0.8 (H12) & 1 & 0 & $0.36\pm0.02$ & $0.17\pm0.02$& 0.70 & 0.26\\
Prolate & 10.5-10.8 & 0.6-0.8 (H12) & 1 & 0 & $0.42\pm0.03$ & $0.20\pm0.02$ & 0.29 & 0.35\\
Prolate & 10.1-10.5 & 0.6-0.8 (H12) & 1 & 0 & $0.45\pm0.02$ & $0.17\pm0.02$ & 0.24 & 0.41\\
\hline
Prolate & 10.8-11.5 & 1-2.5 & 1 & 0 & $0.44\pm0.03$ & $0.21\pm0.02$ &  0.90 & 0.48\\
Prolate & 10.5-10.8 & 1-2.5 & 1 & 0 & $0.45\pm0.02$ & $0.17\pm0.02$ & 0.55 & 0.37\\
Prolate & 10.1-10.5 &  1-2.5 &  1 & 0 & $0.43\pm0.02$ & $0.17\pm0.02$ & 0.61 & 0.47\\
\hline 
\multicolumn{9}{|c|}{Single triaxial model} \\
\hline
Triaxial & 10.8-11.5 &  0.04-0.08 (SDSS) & $0.60^{+0.00}_{-0.08}$ & $0.16^{+0.00}_{-0.12}$ & $0.45^{+0.02}_{-0.00}$ & $0.23^{+0.00}_{-0.01}$ & *0.03* & 0.15\\
Triaxial & 10.5-10.8 & 0.04-0.08 (SDSS) &  $0.92^{+0.00}_{-0.92}$ & $0.00^{+0.66}_{-0.00}$ & $0.47^{+0.16}_{-0.00}$ & $0.24^{+0.00}_{-0.06}$ & *0.00* & 0.12\\
Triaxial & 10.1-10.5 & 0.04-0.06 (SDSS) &  $0.92^{+0.02}_{-0.92}$ & $0.00^{+0.24}_{-0.00}$ & $0.50^{+0.17}_{-0.00}$ & $0.26^{+0.00}_{-0.11}$ & *0.03* & 0.13\\
\hline
Triaxial & 10.8-11.5 & 0.6-0.8 (H12) & $0.76^{+0.00}_{-0.62}$ & $0.92^{+0.00}_{-0.92}$ & $0.44^{+0.06}_{-0.05}$ & $0.20^{+0.05}_{-0.04}$ & 0.82 & 0.34\\
Triaxial & 10.5-10.8 & 0.6-0.8 (H12) & $0.92^{+0.00}_{-0.72}$ & $0.00^{+0.06}_{-0.00}$ & $0.43^{+0.20}_{-0.00}$ & $0.25^{+0.09}_{-0.06}$ & 0.39 & 0.18\\
Triaxial & 10.1-10.5 & 0.6-0.8 (H12)  & $0.92^{+0.04}_{-0.88}$ & $0.00^{+0.28}_{-0.00}$ & $0.47^{+0.18}_{-0.00}$ & $0.21^{+0.04}_{-0.09}$ &  0.40 & 0.26\\
\hline
Triaxial & 10.5-10.8 & 1-2.5 & $0.00^{+0.72}_{-0.00}$ & $0.20^{+0.08}_{-0.20}$ & $0.61^{+0.05}_{-0.10}$ & $0.13^{+0.06}_{-0.05}$ & 0.53 & 0.38\\
Triaxial & 10.8-11.5 &  1-2.5 &  $0.92^{+0.04}_{-0.82}$ & $0.00^{+0.04}_{-0.00}$ & $0.47^{+0.17}_{-0.03}$ & $0.26^{+0.02}_{-0.11}$ & 0.75 & 0.33\\
Triaxial & 10.1-10.5 & 1-2.5 & $0.00^{+0.80}_{-0.00}$ & $0.52^{+0.00}_{-0.52}$ & $0.56^{+0.05}_{-0.07}$ & $0.17^{+0.07}_{-0.03}$ & 0.70 & 0.47\\
\hline
\end{tabular}
\end{table*}

For each of the SDSS (vdW09), COSMOS/GEMS (H12), and CANDELS data sets we search for
the best-fitting triaxial model as described above on a grid spaced as
($\Delta T$, $\Delta \sigma_T$, $\Delta E$, $\Delta \sigma_E$)=(0.04,
0.04, 0.01, 0.01). We separately consider the two special cases:
oblate, with $T=0$ and $\sigma_T=0$; and prolate, with $T=1$ and
$\sigma_T=0$.  As before, the samples are analyzed in bins of stellar
mass and redshift.  The results are given in Table~\ref{model_single}
and a subset are shown in Figure~\ref{ca_fig_triaxial}.  For each best-fitting model we
estimate the goodness-of-fit by computing the K-S and M-W probabilities that the observed $q_{data}$
represent a population of galaxies with a projected axis ratio
distribution $q_{model}$. 
Note that our fitting method does not aim to maximize the
probabilities given by these goodness-of-fit indicators.

As noted by H12, the axis ratio distribution of
present-day early-type galaxies cannot generally be accurately
described by a single-component model with Gaussian distributions for
the intrinsic parameters.  The one exception is that massive
early-type galaxies ($\log(M_*/M_{\odot})>10.8$) quite closely resemble a
single, highly triaxial population ($T=0.6$).
At all redshifts up to $z=2.5$
no prolate model fits the data, while an oblate model cannot be ruled
out.  The oblate model fitting results reflect the previously
mentioned evolution in the median axis ratio: the intrinsic
ellipticity for the most massive galaxies increases from $E=0.48$ at
$z<0.1$ to $E= 0.61$ at $z>1$, while it decreases from $E=0.66$ to
$E=0.56$ for galaxies in the mass range
$10.1<\log(M_*/M_{\odot})<10.5$.

\subsubsection{Two-component Model for the Intrinsic Shapes}
\label{sec5_1_2}
\begin{figure*}[ht]
\centering
\includegraphics[width=1.0\textwidth]{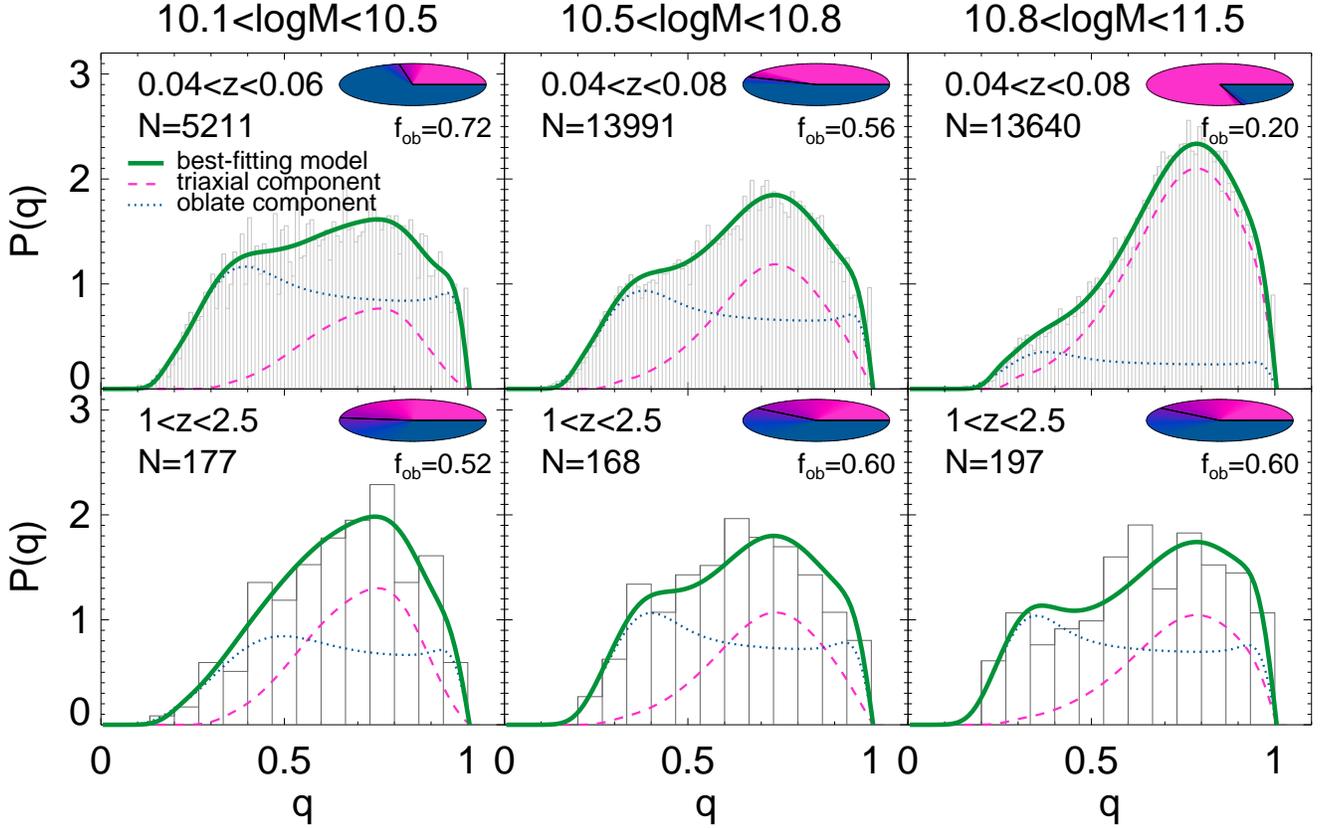}
\caption{Histograms show observed distributions of projected axis ratios for present-day early-type galaxies from SDSS (upper row) and at $1<z<2.5$ from CANDELS (bottom row), each in three mass bins.  The green lines represent the best-fitting, two-component models with Gaussian distributions for intrinsic axis ratios as described in Section~\ref{sec5_1_2}. The dashed pink lines respresent the triaxial component; the dotted blue lines represent the oblate component.  The parameters characterizing the Gaussians are given in Tables~\ref{model_sdss} and \ref{model_two}.  The small pie charts represent $f_{ob}$, the oblate fraction, and its uncertainty.  For the CANDELS sample, the triaxial components are assumed to be identical to the best-fitting triaxial components found for the SDSS sample in the same mass bin.  The strong dependence of the oblate fraction on galaxy mass is much weakened at $z>1$.  The most striking feature is the large fraction of oblate, that is, disk-like galaxies in the high-mass bin.}
\label{ca_fig_par3_pie}
\end{figure*}

Because the single-component models with Gaussian distributions for
the intrinsic shape parameters cannot reproduce the shape distribution
of the low-redshift sample, we now explore a different approach.
As shown most recently by H12, a two-component model can accurately
describe the axis-ratio distribution of present-day early-type
galaxies over a large range in mass.  One of these components is
triaxial, precisely of the form used above; the other component is
oblate, with a normally distributed intrinsic axis ratio, with mean
$b$ and standard deviation $\sigma_b$.  Thus, we now have six parameters
that describe the intrinsic shape distribution; the seventh free parameter
is the fraction assigned to the oblate component ($f_{ob}$).  The
spacing of the grid we now use to search for the best-fitting model is
($\Delta f_{ob}$, $\Delta T$, $\Delta \sigma_T$, $\Delta E$, $\Delta
\sigma_E$, $\Delta b$, $\Delta \sigma_b$)=(0.04, 0.04, 0.02, 0.01,
0.01, 0.01, 0.01, 0.01).

\begin{table*}[ht]
\caption{Double-component Fitting Results for $z=0$}
\label{model_sdss}
\centering
\begin{tabular}{|cc||ccccccc|cc|}
\hline 
Mass & redshift(z) &  $f_{ob}$ \footnote{$f_{\rm{ob}}$ is the fraction of the oblate component.} & $b$ \footnote{$b$ the intrinsic axis ratio of the oblate component and $\sigma_b$ its standard deviation.} & $\sigma_b$ & $T$ \footnote{$T$ is the mean triaxiality parameter, with standard deviation $\sigma_{\rm{T}}$; these are set to 0 or 1 for the oblate and prolate models.}& $\sigma_T$ & $E$ \footnote{$E$ and $\sigma_{E}$ are the ellipticity (1 minus the intrinsic short-long axis ratio) and its standard deviation.}  & $\sigma_E$  & $P_{KS}$ \footnote{The final two columns list the K-S and M-W probabilities that the observed and best-fitting model projected axis ratio distributions are indistinguishable, for a randomly drawn realization of the model distribution with the same number of objects as the observed samples.  These serve as a crude goodness-of-fit test.} & $P_{MW}$ \\
\hline 
10.8-11.5 & 0.04-0.08 (SDSS) & $0.20\pm0.02$ \footnote{Uncertainties are obtained from bootstrapping.} & $0.29\pm0.02$ & $0.07\pm0.01$ & $0.64\pm0.06$ & $0.08\pm0.05$ & $0.41\pm0.02$ & $0.19\pm0.02$& 0.26 & 0.46\\ 
10.5-10.8 & 0.04-0.08 (SDSS) & $0.56\pm0.06$ & $0.28\pm0.01$ & $0.08\pm0.01$ & $0.68\pm0.12$ & $0.08\pm0.06$ & $0.45\pm0.02$ & $0.16\pm0.03$ & 0.29 & 0.19\\ 
10.1-10.5 & 0.04-0.06 (SDSS) & $0.72\pm0.06$ & $0.28\pm0.01$ & $0.09\pm0.01$ & $0.48\pm0.08$ & $0.08\pm0.06$ & $0.49\pm0.02$ & $0.12\pm0.02$  & 0.84 & 0.28 \\ 
\hline
\end{tabular}
\end{table*}

The two-component approach results in a very good description of the
observed axis ratio distributions of present-day galaxies (see Table~\ref{model_sdss}
and Figure~\ref{ca_fig_par3_pie}).  The goodness-of-fit indications from the K-S and M-W
statistical tests suggest that the best-fitting models provide a
realistic view of the intrinsic shape distribution.  Over the entire
galaxy mass range, a highly triaxial ($T\sim 0.6$), yet flattened
($E\sim 0.45$), component combined with an even flatter ($b\sim 0.3$)
oblate component provides a good description of the data, with little
variation in these shape parameters with galaxy mass.  The parameter
that captures the strong mass-dependence in galaxy structure is
$f_{ob}$, the fraction assigned to the second, oblate component: it
rises from $f_{ob} = 0.20\pm 0.02$ at high mass to $f_{ob} = 0.72\pm
0.06$ at low mass.

These results are very similar to those presented by H12 -- small
differences occur due the choice of different stellar mass bins as
well as a different implementation of the intrinsic variation in the
shape parameters -- the $\sigma$ parameters -- in generating the
probability distributions $p(q_{model})$.

\begin{table*}[ht]
\caption{Double-component Fitting Results for $z=0.6-2.5$}
\label{model_two}
\centering
\begin{tabular}{|cc||ccc|cc||c|cc|}
\hline
\multirow{2}{*}{Mass}  & \multirow{2}{*}{Redshift(z)} 
& \multicolumn{5}{c||}{Oblate Parameters Free \footnote{Fix triaxial component in the same stellar mass bins as local galaxies.} } & 
\multicolumn{3}{c|}{Oblate Fraction Free \footnote{Fix other parameters in the same stellar mass bins as local galaxies.} } 
 \\
 \cline{3-10}
 & &  $f_{ob}$ \footnote{$f_{\rm{ob}}$ is the fraction of the oblate component.} & $b$ \footnote{$b$ the intrinsic axis ratio of the oblate component and $\sigma_b$ its standard deviation.} & $\sigma_b$  & $P_{KS}$ \footnote{The final two columns list the K-S and M-W probabilities that the observed and best-fitting model projected axis ratio distributions are indistinguishable, for a randomly drawn realization of the model distribution with the same number of objects as the observed samples.  These serve as a crude goodness-of-fit test.} & $P_{MW}$  &  $f_{ob}$  & $P_{KS}$ & $P_{MW}$ \\
\hline
10.8-11.5 & 0.6-0.8 (H12) & $0.16\pm0.18$  \footnote{Uncertainties are obtained from bootstrapping.} & $0.33\pm0.10$ & $0.05\pm0.08$  & 0.70 & 0.36 & $0.12\pm0.06$& 0.43 & 0.26 \\
10.5-10.8 & 0.6-0.8 (H12) & $0.48\pm0.23$ & $0.28\pm0.07$ & $0.05\pm0.06$ &  0.21 & 0.24&  $0.49\pm0.08$& 0.22 & 0.24\\
10.1-10.5 & 0.6-0.8 (H12) & $0.56\pm0.12$ &  $0.32\pm0.03$ & $0.06\pm0.04$ & 0.87 & 0.36 &  $0.51\pm0.08$& 0.67 & 0.38\\
\hline
10.8-11.5 & 1-2.5 & $0.60\pm0.24$ & $0.27\pm0.07$ & $0.06\pm0.05$  & 0.71 & 0.49 & $0.59\pm0.10$& 0.54 & 0.40\\
10.5-10.8 & 1-2.5 & $0.60\pm0.25$ & $0.31\pm0.07$ & $0.07\pm0.03$ & 0.87 & 0.37 & $0.53\pm0.14$& 0.61 & 0.33\\
10.1-10.5 & 1-2.5 & $0.52\pm0.24$ & $0.34\pm0.10$ & $0.12\pm0.06$  & 0.69 & 0.46 & $0.38\pm0.11$& 0.12 & 0.16\\
\hline
\multicolumn{10}{|c|}{Redshift Bins of CANDELS} \\
\hline
10.8-11.5 & 0.6-0.8 & $1.00\pm0.31$ & $0.52\pm0.13$ & $0.25\pm0.10$  & 0.96 & 0.44 & $0.35\pm0.20$ & 0.25 & 0.29\\
10.8-11.5 & 0.8-1.3 & $0.84\pm0.21$ & $0.29\pm0.05$ & $0.06\pm0.05$  & 0.99 & 0.50 & $0.81\pm0.20$ & 0.97 & 0.44\\
10.8-11.5 & 1.3-1.8 & $0.48\pm0.28$ & $0.22\pm0.08$ & $0.05\pm0.07$  & 0.89 & 0.46 & $0.59\pm0.16$ & 0.73 & 0.48\\
10.8-11.5 & 1.8-2.5 & $1.00\pm0.27$ & $0.41\pm0.09$ & $0.20\pm0.07$  & 0.97 & 0.44 & $0.51\pm0.21$ & 0.94 & 0.45\\
\hline
10.5-10.8 & 0.6-0.8 & $0.36\pm0.31$ & $0.27\pm0.10$ & $0.05\pm0.09$ & 0.92& 0.46 & $0.42\pm0.24$ & 0.89 & 0.49\\
10.5-10.8 & 0.8-1.3 & $1.00\pm0.28$ & $0.40\pm0.07$ & $0.16\pm0.07$ & 0.99 & 0.49 & $0.53\pm0.27$ & 0.87 & 0.48\\
10.5-10.8 & 1.3-1.8 & $0.64\pm0.31$ & $0.30\pm0.10$ & $0.05\pm0.08$ & 0.67 & 0.30 & $0.63\pm0.24$ & 0.66 & 0.31\\
10.5-10.8 & 1.8-2.5 & $0.56\pm0.32$ & $0.29\pm0.09$ & $0.05\pm0.03$ & 0.93 & 0.39 & $0.55\pm0.28$ & 0.92 & 0.38\\
\hline
10.1-10.5 & 0.6-0.8 & $0.56\pm0.28$ & $0.26\pm0.09$ & $0.05\pm0.08$ & 0.63 & 0.32 & $0.57\pm0.19$ & 0.68 & 0.37\\
10.1-10.5 & 0.8-1.3 & $0.84\pm0.26$ & $0.46\pm0.08$ & $0.17\pm0.10$ & 0.97 & 0.45 & $0.31\pm0.18$ & 0.41 & 0.28\\
10.1-10.5 & 1.3-1.8 & $0.44\pm0.36$ & $0.23\pm0.12$ & $0.19\pm0.10$  & 0.37 & 0.19 & $0.47\pm0.20$ & 0.94 & 0.42\\
10.1-10.5 & 1.8-2.5 & $0.80\pm0.28$ & $0.31\pm0.10$ & $0.05\pm0.06$  & 0.47 & 0.28 & $0.71\pm0.35$ & 0.82 & 0.46\\
\hline
\end{tabular}
\end{table*}

The high-redshift samples are too small to be treated with seven 
independent free parameters. However, given the success of the
two-component model in describing the shape distribution of
present-day early-type galaxies, we can use our superior knowledge of
the low-redshift population to inform the model for the high-redshift
population. Because each of the two components are very similar across
the mass range explored here for the low-redshift sample, we assume
that the same components can be used as an appropriate model to
describe the higher-redshift observations.  First, we use the
best-fitting triaxial component for each of the three mass bins, with
fixed intrinsic shape distributions, but let the oblate component vary
arbitrarily.  That is, the parameters $b$, $\sigma_b$, and $f_{ob}$ are
allowed to vary, while the others are kept fixed.  The results are
shown in Table~\ref{model_two} and Figure~\ref{ca_fig_par3_pie}.

For the COSMOS+GEMS and combined ($1<z<2.5$) CANDELS samples we find
that all evolution with redshift can be accounted for by evolution in
$f_{ob}$; no significant changes in $b$ (or $\sigma_b$) are seen.  For
the highest-mass galaxies ($10.8<\log(M_*/M_{\odot})<11.5$) $f_{ob}$ is seen to rise at $z>1$, from
$f_{ob}\sim 0.2$ at $z<1$ to $f_{ob}=0.60\pm 0.24$.  The large
uncertainty is due to the degeneracy between $b$ and $f_{ob}$:
evolution in the average shape can either be accommodated by a change
in the average shape of the galaxies represented by the oblate
component, or by a change in the fraction of oblate galaxies.  The
unsubstantial changes in $b$ ($\sigma_b$) with mass and redshift
motivate us to implement a second restriction to our model: we now
keep all intrinsic shape parameters at the values found for the
low-$z$ SDSS sample, and only allow $f_{ob}$ to vary.  

\begin{figure}[ht]
\centering
\includegraphics[width=1.0\columnwidth]{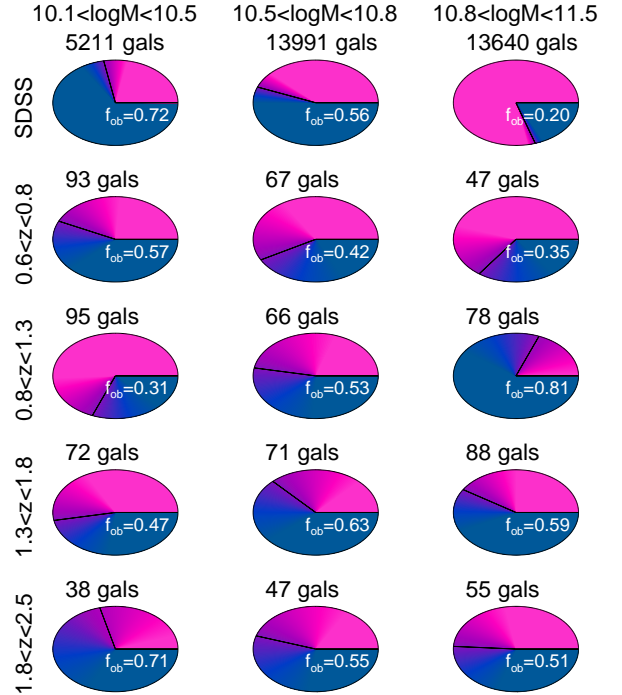}
\caption{The evolution of the oblate fraction of early-type galaxies in different mass bins.  Compared to the results shown in Figure~\ref{ca_fig_par3_pie}, the redshift bins at $z>1$ are now narrower, as indicated, and the intrinsic shape distributions of both the oblate and the triaxial component are kept fixed at the values found for the present-day SDSS sample.  The values of $f_{ob}$ and their uncertainties are given in Table~\ref{model_two}.}
\label{ca_fig_pie}
\end{figure}

This restriction seems justified by the results from the
goodness-of-fit tests: the predicted distribution from the
best-fitting models, even with only a single free parameter
($f_{ob}$), do not significantly differ from the observed
distributions according the the the K-S and M-W tests.  
The results are also shown in Table~\ref{model_two} and Figure~\ref{ca_fig_pie}.
We now find that the oblate fraction for the massive galaxies increases from $0.20\pm0.02$
at $z<0.1$ to $0.59\pm0.10$ at $z>1$, a highly significant (4$\sigma$)
change. For galaxies in our middle mass bin ($10.5<\log(M_*/M_{\odot})<11.8$), $f_{ob}$ does not
change with redshift and stays at $\sim 0.5-0.6$, whereas, remarkably,
$f_{ob}$ significantly declines from $0.72\pm0.06$ to $0.38\pm0.11$
for low-mass galaxies ($10.1<\log(M_*/M_{\odot})<10.5$).  The latter was already reflected by the
increased median axis ratio with redshift (see Section~\ref{sec3}).

\section{Discussion}
\label{sec6}
$L^*$ early-type galaxies ($M_*\sim 10^{11}M_\odot$) in the
present-day universe possess a wide range of intrinsic
  shapes: there is no single oblate, prolate, or
triaxial shape that, viewed from any number of random viewing
angles, can account for their projected axis ratio distribution
\citep[e.g., ][]{1992MNRAS.258..404L, 1996AJ....111.2243T}.  We
implemented two methods to describe and model this distribution.
First, we showed that a single family of oblate or prolate structures
with broadly distributed intrinsic axis ratios accurately captures the
observed projected distribution (Section~\ref{sec4} and
Figure~\ref{ca_fig_deprojections}).  Second, we showed that a
combination of triaxial and oblate structures, with normally
distributed intrinsic shapes, works equally well.
This second approach is attractive as the distinction of two components corresponds to the kinematical distinction between `fast rotators' and `slow rotators' \citep[e.g.,][]{2011MNRAS.414..888E}.
Figure~\ref{ca_fig_par3_pie} shows that a triaxial component combined
with a thinner, oblate component provides a good description over a
large range of galaxy masses.  The strong dependence of galaxy
structure on stellar mass is driven by the variation in the relative
abundances of triaxial and oblate objects.  We now discuss the
evolution of the intrinsic shape distribution of early-type galaxies,
based on our analysis presented in Section~\ref{sec3}, \ref{sec4}, and \ref{sec5}.

\subsection{Increased Incidence of Disk-Like, Massive Early-type Galaxies at $z>1$}
\label{sec6_1}

\begin{figure}[ht]
\centering
\includegraphics[width=1.0\columnwidth]{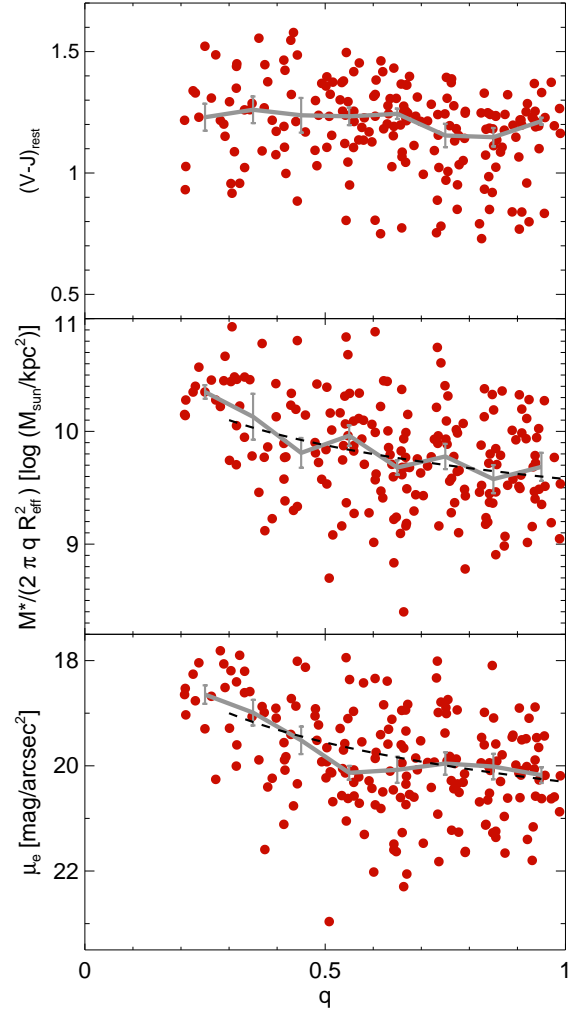}
\caption{The projected axis ratio vs. rest-frame $V-J$ color, mass surface density, and surface brightness for the early-type galaxies selected from CANDELS at redshifts $1<z<2.5$ and more massive than $\log(M_*/M_\odot)>10.8$. The gray lines with error bars from bootstrapping represent running medians.  The lack of a trend in the top panel suggests that these galaxies contain little or no dust; otherwise, galaxies with small axis ratios, that is, those viewed edge-on, would be expected to have redder colors.  The increased surface mass density and surface brightness of the small-axis ratio galaxies (bottom two panels) suggest that these galaxies are oblate rather than prolate; the dashed black lines are the expected projected surface brightness and density for an oblate model, with intrinsic axis ratio $E=0.61$, which is the best-fitting value from the single-component model for this sample (Table~\ref{model_single}).}
\label{ca_fig_dust}
\end{figure}

The cumulative distributions of projected axis ratios of $L^*$
early-type galaxies at $z>1$ and at the present-day show that these
were on average flatter in the past (Section~\ref{sec3}; Figure~\ref{ca_fig_q_hist}).  Our
parameterized modeling approach presented Section~\ref{sec5_1_2} 
interprets this as a change in the fraction of the oblate component,
from $0.20\pm0.02$ at $z<0.1$ to $0.59\pm0.10$ at $z>1$.

Because the $z>1$ sample is too small to directly distinguish what
structural family the galaxies belong to, we consider independent
evidence for our interpretation that the $z>1$ population largely
consists of flat, oblate objects.  At the present day, flatness is
associated with rotation \citep[e.g.,
][]{2008MNRAS.387...79V,2011MNRAS.414..888E}, but so far such
kinematic evidence has not been extended beyond $z\sim1$
\citep{2008ApJ...684..260V}.  The best direct evidence for our
interpretation that flat galaxies in our sample are indeed disk-like
in structure is that the stellar surface mass density (middle panel of
Figure~\ref{ca_fig_dust}) and the surface brightness (bottom panel of
Figure~\ref{ca_fig_dust}) are larger for galaxies with small projected
axis ratios. This is expected in case the flat galaxies are edge-on
and oblate, but not if they are edge-on and prolate. In the latter
case, the flattest galaxies should have the smallest surface
brightness.  We note that these considerations are only valid for
transparent, that is, dust-poor, stellar systems. This 
assumption is supported by the observation that the rest-frame
$V-J$ color does not significantly change with projected axis ratio,
implying little variation in dust attenuation with inclination and,
thus, a low dust content. The lack of star formation activity in
these objects combined with their low dust content indicate that our
sample consists of galaxies with smooth light profiles, and is
therefore comparable to a morphologically classified sample of
early-type galaxies based on visual inspection of images.

Further direct evidence of prominent disks in high-redshift early-type
galaxies comes from two-dimensional bulge-disk decompositions
\citep{2006ApJ...650..706S,2008ApJ...672..146S,2008ApJ...682..303M,2011ApJ...730...38V,2012MNRAS.427.1666B}

Based on these independent lines of evidence, we conclude that at
$z>1$ a substantially larger fraction of $L^*$ early-type galaxies are
disk-like than at $z<1$.  This evolution in structure coincides with
evolution in size
\citep[e.g.,][]{2007ApJ...656...66Z,2007ApJ...671..285T,2008ApJ...677L...5V,
  2008ApJ...688...48V,2012ApJ...746..162N}.  Van der Wel et
al. (2013a, in preparation) showed that the number density of small
($\lesssim 2$kpc) early-type galaxies dramatically decreases between
$z\sim 2$ and the present day \citep[also see][]{,2011ApJ...743...96C,2013arXiv1303.2689C}.
These early types are, as we have shown here, commonly disk-like, such
that we may conclude that individual galaxies evolve from small and
disk-like at $z\sim 2$ to large and round at $z\sim 0$.  The evolution
of size and internal structure  could be driven by a single process, 
and merging is usually considered to be the most plausible process \citep[e.g., ][]{2010ApJ...719..844R,2012ApJ...744...85M,2012ApJ...746..162N}. 
Major merging and more smooth growth in mass
through accretion and disruption of satellites can account for the
disappearance of prominent disks in $L^*$ early types at $z\sim 2$,
and the observation that the most massive galaxies in the present-day
universe do not host disks (vdW09).

In addition to the growth of individual galaxies, evolution in the
population is driven by the strong increase in the number density of
early-type galaxies between $z\sim 2$ and the present
\citep[e.g.,][]{2004ApJ...608..752B,2007ApJ...665..265F,2011ApJ...739...24B}.
From $z=2$ to $z=1$ the fraction of triaxial galaxies increases, but
this cannot be the result of the formation of `new' early-type
galaxies in the form of triaxial systems from 
already-formed early-type galaxies. The absolute number densities
of oblate and triaxial systems both increase over that time span, and we
suggest that all `new' early types start out as compact and disk-like
and subsequently evolve into larger, more triaxial systems
\citep[e.g., ][]{2009ApJ...699L.178N,2012ApJ...744...63O}.  This
suggestion is motivated by the notion that the immediate progenitors
of `new' early-type galaxies will be gas rich and star-forming,
creating suitable circumstances for the formation of disks (see
below), and by the notion that it is implausible that round, triaxial
systems evolve into disk-like systems in the absence of star
formation.  At $z<1$ a natural balance is established between the
addition of `new', disk-like early types and the gradual formation of
triaxial systems, resulting in an almost unchanging, but strongly
varied mix of intrinsic structures, as discussed by H12.

In the scenario described above, galaxies in which star formation is
truncated retain the disk-like structure of their presumed,
star-forming progenitors.  A full discussion of the transition process
is beyond the scope of this paper, but it is important to point out
that while newly formed early-type galaxies retain disk-like
properties, their light (and stellar mass) distributions are more
centrally concentrated than those of equally massive star-forming
galaxies \citep[e.g.,][]{2009ApJ...705..255T, 2011ApJ...742...96W, 2012ApJ...753..167B}.  
This implies that a substantial increase in the central
stellar density occurs before or at the time of transition.  A
centrally concentrated starburst fueled by a gas-rich merger is one
possible mechanism to produce bulge-like bodies \citep[e.g.,][]{2006ApJ...648L..21K}.  
More recently, violent disk instabilities in a gas-rich
galaxy have been argued to produce clumps that may migrate to the
center on a short time scale, quickly creating a dense stellar body
\citep{2009ApJ...703..785D,2010MNRAS.404.2151C, 2012ApJ...757..120G}.
Whether the gas content of the resulting, dense, disk-like, but
non-star-forming, galaxy has been heated and removed \citep[e.g.,][]{2008ApJS..175..390H} or merely stabilized \citep[e.g.,][]{2009ApJ...707..250M,2012MNRAS.420.3490C} is still debated.

\subsection{Decreased Incidence of Disk-like, Low-mass Early-type Galaxies at $z>1$}
\label{sec6_2}
Sub-$L^*$ early-type galaxies ($M_*\sim 10^{10}M_\odot$) in the local
universe are most often oblate and disk-like.  The comparison with the
cumulative axis ratio distributions of such objects at $z>1$ tells us
that these were less disk-like (see Figure~\ref{ca_fig_q_hist}).  This
may appear to be at odds with the results discussed above,
that is, that massive early types were more disk-like at $z>1$.

Our interpretation of this 3$\sigma$ effect remains largely speculative.
Low-mass early-type galaxies in the present day can be surmised to be
disk-like for the simple reason that their star-forming progenitors
are also disk-like.  Star formation may stop either due to some
internal process or due to environmental effects such as ram-pressure
stripping.  In the latter case the structure of the stellar disk will
remain intact, leading to a very flat early-type galaxy.  At $z\sim 2$
the fraction of satellite galaxies in this mass range is predicted to
be negligible, whereas among the present-day population satellite
galaxies make up $30\%-40\%$ of the total
\citep[e.g.,][]{2008MNRAS.387...79V}. Indeed, the axis ratio distributions
of present-day centrals and satellites are significantly different
\citep{2010ApJ...714.1779V}, but even the present-day centrals are not as
round as their $z>1$ counterparts \citep[also see][]{2011MNRAS.413..921V},.  We suggest that the low-mass
early-type galaxies at $z>1$ are not very disk-like, simply because
their star-forming progenitors were not disk-like at that epoch:
van der Wel et al. (2013b, in preparation) showed that low-mass ($M_*<M_\odot^{10}$) star-forming galaxies
at $z>1$ had not yet attained stable, rotating structures, like they
have at later epochs.  Whether this is related remains to be seen and
hinges on our general lack of understanding of how star-forming galaxies
are transformed into passive, early-type galaxies.

\section{Summary}
\label{sec7}
Projected axis ratio measurements from HST/WFC3 F160W imaging from CANDELS of 880 early-type galaxies at redshifts $1<z<2.5$, complete down to a stellar mass of $\log(M_*/M_{\odot}) = 10$ over the whole redshift range, are used to reconstruct and model their intrinsic shapes.  The sample is selected by low star-formation activity on the basis of $U-V$ and $V-J$ rest-frame colors (see Figure \ref{ca_fig_uvj}, and we demonstrate that these galaxies are dust-poor and transparent: those with flat projected shapes have the same colors as those with round shapes (see Figure~\ref {ca_fig_dust}, top panel).  In addition, the increased surface mass densities and surface brightness of flat galaxies (Figure~\ref{ca_fig_dust}, bottom two panels) suggest that flattening is associated with a disk-like internal structure; prolate galaxies would have lower surface densities when viewed edge-on. Therefore, we conclude that our sample consists of genuine early types, comparable to samples based on visual morphological classification.  We compare the shape distribution of this sample with the shape distribution of early-type at low redshift ($0.04 < z < 0.08$) selected in a similar manner from the SDSS.

Similar to their present-day counterparts, the $z>1$ early-type galaxies show a large variety in intrinsic shape; even at a fixed mass, the projected axis ratio distributions cannot be explained by random projection of a set of galaxies with very similar intrinsic shapes.  We demonstrated this in two ways by assuming that all galaxies are oblate (or prolate): first, an analytical approximation to deproject the observed axis ratio distributions implies that a very broad range in intrinsic shapes is required (Section~\ref{sec4} and Figure~\ref{ca_fig_deprojections}); second, we show that randomly projecting a set of objects with a Gaussian distribution of intrinsic axis ratios cannot match the observed, projected shape distribution (Section~\ref{sec5_1_1} and Figure~\ref{ca_fig_triaxial}).

As was demonstrated for present-day early-type galaxies and up to $z\sim 1$, a two-population model can accurately describe the projected axis ratio distributions. We now extend this to $z=2.5$.  This model, inferred from fitting the axis ratio distribution of the low-redshift sample (Section\ref{sec5_1_2} and Figure~\ref{ca_fig_par3_pie}), consists of a triaxial, fairly round population combined with a flat ($c/a\sim 0.3$) oblate population.  For present-day early-type galaxies the oblate fraction strongly depends on galaxy mass, but at $z>1$ this trend is not seen over the stellar mass range explored here ($10<\log (M_*/M_{\odot})<11.3$).  This is mostly the result of strong evolution in the oblate fraction among high-mass early-type galaxies: for galaxies with mass $\log(M_*/M_{\odot})>10.8$ the oblate fraction increases from $0.20\pm 0.02$ at the present day to $0.59\pm 0.10$ at $1<z<2.5$.  Conversely, we find that the oblate fraction decreases with redshift for low-mass early-type galaxies ($\log (M_*/M_{\odot})<10.5$), from $0.72\pm 0.06$ to $0.38\pm 0.11$. These results are based on the assumption that the intrinsic shapes of the triaxial and oblate population do not evolve with redshift. We refer to Section~\ref{sec5_1_2} for a justification of this assumption and a demonstration that our results and interpretation do not depend on it.

The decreased prevalence of disk-like systems and larger galaxy sizes at lower redshifts point to a scenario in which classical elliptical galaxies gradually emerge over time through merging and the accretion of satellites, at the expense of the destruction of pre-existing disks.  Definitive evidence for the disk-like structure of massive early-type galaxies at $z\sim 2$ should eventually be provided by kinematic evidence for rotation in the stellar body. We speculate that the decreased incidence of disks at early cosmic times among low-mass early-type galaxies can be attributed to two factors: low-mass, star-forming progenitors at $z>1$ were not settled into stable disks to the same degree as at later cosmic times, and the stripping of gas from satellite galaxies is an increasingly important process at lower redshifts.  We refer to Section~\ref{sec6_2} for a discussion.

A joint analysis of shapes, sizes, and S\'ersic indices for late- and early-type galaxies, will provide further insight into the intrinsic structure of high-redshift galaxies, and allow for more constrained deprojection and model construction approaches.  Further improvements will be provided by the extension of the analysis to the full CANDELS data set, drawing samples from all five  fields instead of the two fields used here; at the moment we are still limited by small number statistics at $z\sim 2$ and above.

\acknowledgments
We thank the anonymous referee for helpful comments, 
and Steve Willner and Matthew Ashby for useful suggestions.
This work is based on observations taken by the CANDELS Multi-Cycle Treasury Program with the NASA/ESA HST, which is operated by the Association of Universities for Research in Astronomy, Inc., under NASA contract NAS5-26555.
Y.-Y. C. was funded by the IMPRS for Astronomy \& Cosmic Physics at the University of Heidelberg and the Marie Curie Initial Training Network ELIXIR of the European Commission under contract PITN-GA-2008-214227.

\bibliographystyle{apj}

\end{document}